# Self-referencing 3D characterization of ultrafast optical-vortex beams using tilted interference TERMITES technique

Jinyu Pan,[1,2,3,†] Yifei Chen,[1,†] Zhiyuan Huang,[1,*] Cheng Zhang,[3] Tiandao Chen,[1,2] Donghan Liu,[1,2] Ding Wang,[1] Meng Pang,[1,3,*] and Yuxin Leng[1,3,*]

[1]*State Key Laboratory of High Field Laser Physics and CAS Center for Excellence in Ultra-intense Laser Science, Shanghai Institute of Optics and Fine Mechanics, Chinese Academy of Sciences, Shanghai 201800, China*

[2]*Center of Materials Science and Optoelectronics Engineering, University of Chinese Academy of Sciences, Beijing 100049, China*

[3]*Hangzhou Institute for Advanced Study, University of Chinese Academy of Sciences, Hangzhou 310024, China*

[*]*E-mail: huangzhiyuan@siom.ac.cn; pangmeng@siom.ac.cn; lengyuxin@siom.ac.cn*

[†]*These authors contributed equally to this work.*

**Abstract:** Femtosecond light pulses carrying optical angular momentums (OAMs), possessing intriguing properties of helical phase fronts and ultrafast temporal profiles, enable many applications in nonlinear optics, strong-field physics and laser micro-machining. While generation of OAM-carrying ultrafast pulses and their interactions with matters have been intensively studied in experiments, three-dimensional characterization of ultrafast OAM-carrying light beams in spatio-temporal domain has, however, proved difficult to achieve. Conventional measurement schemes rely on the use of a reference pulsed light beam which needs to be well-characterized in its phase front and to have sufficient overlap and coherence with the beam under test, largely limiting practical applications of these schemes. Here we demonstrate a self-referencing set-up based on a tilted interferometer that can be used to measure complete spatio-temporal information of OAM-carrying femtosecond pulses with different topological charges. Through scanning one interferometer arm, the spectral phase over the pulse spatial profile can be obtained using the tilted interference signal, and the temporal envelope of the light field at one particular position around its phase singularity can be retrieved simultaneously, enabling three-dimensional beam reconstruction. This self-referencing technique, capable of measuring spatio-temporal ultrafast optical-vortex beams, may find many applications in fields of nonlinear optics and light-matter interactions.

**Keywords:** ultrafast pulse measurement, optical vortex, optical angular momentum, self-referencing measurement, spatio-temporal characterization

## 1. Introduction

Optical-vortex beams carrying orbital angular momentums (OAMs), with phase singularities in their centers and hollow-structured intensity distributions,[1-3] have proved important in many areas of optics and physics.[4-7] In the past several decades, intensive studies on the helical phase fronts of optical-vortex beams have deepened our fundamental understanding on light fields, and meanwhile the unique phase structures of optical-vortex beams, carrying OAMs, have enabled wide-ranging applications in micro-particle manipulation,[8-10] optical microscopy,[11,12] quantum optics[13-15] and optical telecommunications.[16-18] Due to their great application potential, ultrafast light fields carrying OAMs, being powerful tools in studies of nonlinear optics and light-matter interactions, have also attracted extensive interest, adding a major new dimension in nonlinear laser-frequency conversion,[19-21] high-harmonic generation,[22,23] laser micro-machining,[24-26] and strong-field physics.[27,28]

In the field of nonlinear optics, self-referencing measurements of ultrafast light fields are one of the most important techniques and have been widely used in many experiments.[29-33] Fast and accurate characterization of ultrafast pulses in three dimensions, in one side, is crucial for studies on ultrafast lasers, providing useful information on the key performance of ultrafast light sources.[34-36] In the other side, accurate measurements of ultrafast light fields generated from optical experiments, are essential for understanding the mechanisms of the ongoing nonlinear-optics processes.[37,38] While several self-referencing techniques, such as frequency-resolved optical gating (FROG) and spectral phase interferometry for direct electric-field reconstruction (SPIDER) set-ups,[39,40] have been successfully used to measure spectra-temporal information of ultrafast pulses,[29-32,35,37] advanced self-referencing scheme that is capable of reconstructing ultrafast optical-vortex beams in both spatial and temporal dimensions, has however escaped experimental realization. In previous set-ups for measuring ultrafast optical-vortex pulses,[41-44] one reference beam was generally needed to characterize the phase front of the light beam under test using interferometry. In such measurements, however, the reference beam needs to be fully coherent with the one under test and the two beams should have sufficient overlaps in all the spatial, temporal and spectral dimensions.[41-44] Additionally, the spatial-phase information of the reference beam should be well known before measurements in order to retrieve precisely the phase front of the beam under test.[41-44] All these requirements restrict, to some extent, the practical application of these schemes.

More recently, a novel total electrical-field reconstruction using a Michelson interferometer temporal scan (TERMITES) technique was developed, enabling the spatio-temporal characterization of high-energy femtosecond laser pulses in a self-referencing manner.[35] However, in previous studies the application of this TERMITES technique for reconstructing ultrafast optical-vortex beam with relatively-complex phase and amplitude distributions is still challenging.[41-44] First, the radial-shearing interferometer in conventional TERMITES set-up is not applicable of measuring hollow-structured optical-vortex beams, since the dark centroid of the vortex beam, i.e., phase singularity, can no longer be used as the reference point of the shearing interference. Second, the relatively-large beam size and complicated focusing condition of optical-vortex pulses cause some difficulties in performing high-signal-to-noise-ratio FROG measurements. In this paper, we demonstrate a novel TERMITES set-up in which a tilted interferometer is used to retrieve the spatial phase information of hollow-structured vortex beams, and a FROG set-up based on a type-II crystal to measure the pulse temporal profile with a high signal-to-noise ratio. Because of these two improvements, we achieved for the first time, to the best of our knowledge, self-referenced reconstruction of OAM-carrying ultrafast pulses with topological charges of $|l|$ = 1, 2 and 3. The retrieved results show striking agreement with the numerical simulations, highlighting the great application potential of this scheme in studies of nonlinear optics and strong-field physics concerning ultrafast vortex light.[27,28,45,46]

## 2. Measurement set-up

The modified TERMITES set-up we built up in the experiment is illustrated in **Figure 1**a. As shown in Figure 1a, when an optical-vortex pulse with a linear polarization state was launched into the system, it was divided into two parts by a 50:50 beam splitter (BS1). One part of the pulse was reflected by a convex mirror (CM1), leading to a slightly divergent light beam (plotted as the pink beam in Figure 1a). The other part of the pulse (plotted as the green beam in Figure 1a) passed through a birefringent quartz crystal (BQC), and then was reflected by the scanning mirror (SM) mounted on a piezoelectrical moving stage. The principal axis of the BQC can be adjusted to slightly deviate from the polarization direction of the incident pulse, resulting in the generation of two adjacent pulses with orthogonal polarization states (see Figure 1b). The intensity ratio of the two pulses depends on the deviation angle, while the pulse separation can be controlled through varying the thickness of the BQC. The two reflected light beams were recombined at the BS1, and the output beam were divided into two optical paths at the second 50:50 beam splitter (BS2), see Figure 1a. For the first path that passed through BS2, the interference signal between the two pulses (reflected by CM1 and SM) with the same polarization states was measured using the CCD camera. In the second path, the light beam was focused by a concave mirror (CM2), and a type-II BBO crystal was placed near the focusing point of the beam. Sum-frequency process between the two pulses at different polarization states can be obtained due to the type-II phase-matching, and the resulting sum-frequency signal at shorter wavelengths was filtered out by an optical filter and then recorded using a fiber-coupled spectrometer for the FROG measurement, see Figure 1a.

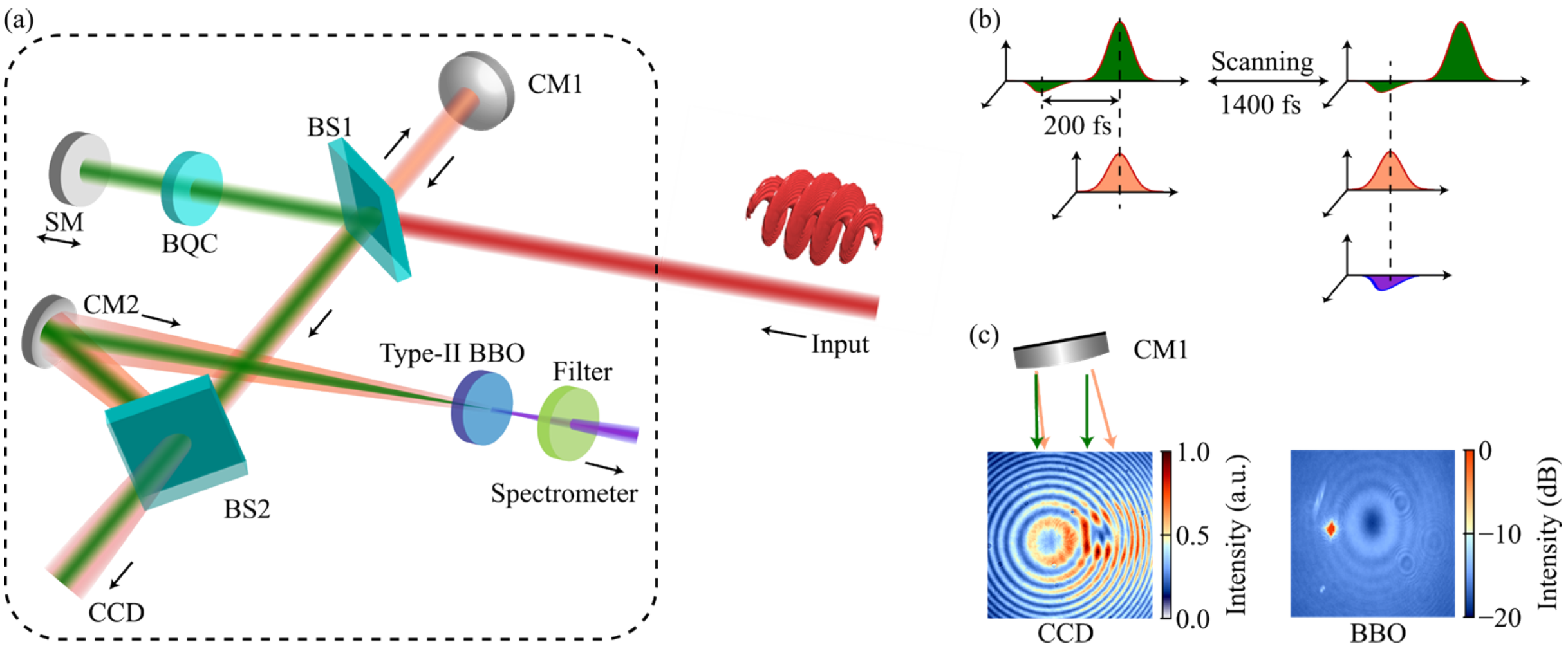

**Figure 1.** a) Schematic of the diagnosis set-up. BS, beam splitter; CM1, convex mirror; BQC, birefringent quartz crystal; SM, scanning mirror; CM2, concave mirror. b) For the green beam, two pulses with orthogonal polarization states are generated with a pulse separation of 200 fs, and the delay scanning range is 1400 fs. c) The tilted shearing-interference signal is recorded by the CCD camera and the sampling point of the FROG measurement locates on one bright ring of the structured light beam.

In the experiment, while the SM was adjusted to be perpendicular to incident light beam, the CM1 was rotated to slightly deviate from the perpendicular position, leading to a tilted shearing-interference measured at the CCD camera (see Figure 1c). This rotation is necessary for measuring optical-vortex beams due to the following two reasons. First, it can lead to the centroid off-set of the two interfering beams at the CCD camera, mitigating the signal fading effect due to dark centers of the beams. Second, the rotation can also result in a sampling point off-set in the BBO crystal over the sum-frequency process (see Figure 1c). Due to the convex surface of CM1, the divergent light beam (marked as pink in Figure 1a) has a longer focusing length than the beam reflected by the flat SM (marked as green in Figure 1a) after the CM2. When the BBO crystal is placed near the focusing point of the green beam, this focusing point works as a sampling spot on the spatial profile of the pink beam (see Figure 1c). In the experiment, the rotation angle of CM1 was carefully adjusted so as to locate the focusing point of the green beam at one bright ring of the pink beam, ensuring a high signal-to-noise ratio of the FROG measurement.

In order to improve the accuracy of the measurement, the axis-angle and thickness of the BQC need to be carefully selected (see Figure 1c). In the experiment, the angle deviation of the BQC principal axis from the polarization state of the incident pulse was ~3%, leading to a portion of merely 0.3% light energy coupled to the orthogonal polarization state for FROG measurement, and this portion of light was focused on the BBO crystal for sampling (see Figure 1b). The almost-equal light intensities of the two pulses at the focusing point ensure that the sum-frequency signal output from the crystal has a high signal-to-noise ratio, suppressing the self-frequency-multiplication signal generated by each of the beams. The rest light beam with 99.7% energy was used for the tilted shearing-interference measurement, resulting in a high interferometric contrast as shown in Figure 1c. An 8-mm-thick BQC was used in the set-up, giving a time delay of 200 fs which is long enough for measuring pulses with a width of <100 fs. The scanning length of the SM is 210 μm (1400 fs scanning delay), sufficient for both spatial phase and FROG measurements.

### 3. Generation of ultrafast light beams carrying OAMs

In order to prepare ultrafast optical-vortex beams for testing the capability of this diagnosis set-up, we used a commercial Ti:Sapphire laser system (Legend Elite, Coherent Inc.) as the seed laser which can deliver linearly-polarized ultrafast light pulses with 1 kHz pulse repetition rate, ~45 fs pulse duration and 790 nm central wavelength. The laser output was collimated to have a Gaussian-shaped profile with a full-width-half-maximum bandwidth of ~9 mm. As shown in **Figure 2**, this Gaussian light beam passed through in turn a quarter wave plate (QWP), a vortex retarder (VR) and a wire grid polarizer (WGP), resulting in the generation of linearly-polarized, ultrafast light beams carrying OAMs.[47,48] The topological charges of optical-vortex pulses can be altered through using different VRs (see Figure S1, Supporting Information).

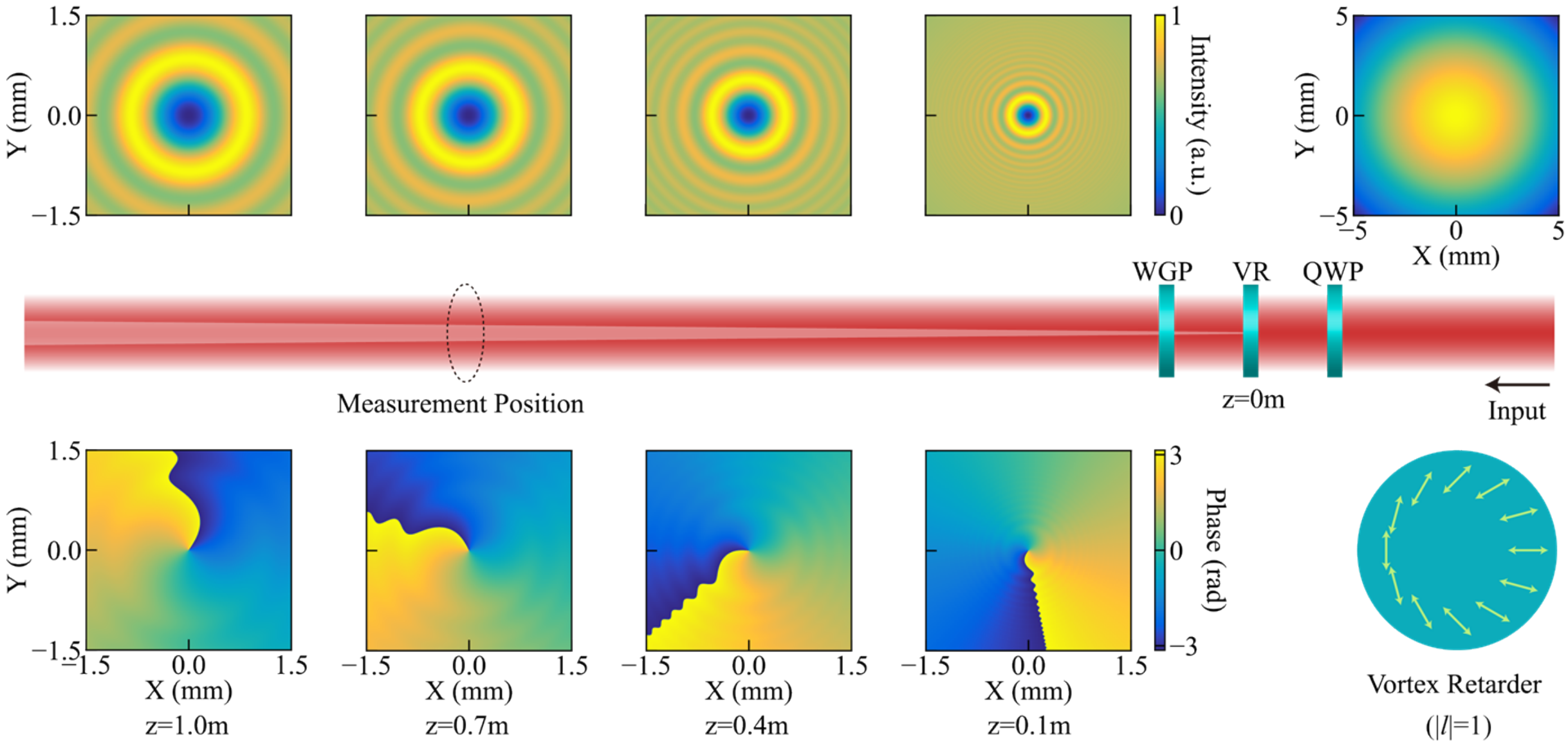

**Figure 2.** The evolution of the light beam over free-space propagation, simulated using Collins formula. A linearly-polarized, Gaussian-shaped light beam passes through in turn a quarter-wave plate (QWP), a vortex retarder (VR) and a wire grid polarizer (WGP). The intensity and phase profiles of the simulated OAM-carrying light beam ($|l| = 1$) at different positions behind the vortex retarder are presented in this plot.

Using Collins formula,[49,50] we simulated theoretically the light evolution over free-space propagation, unveiling the formation process of phase singularity in the centroid of the light beam. The simulated results of the optical-vortex beam with a topological charge of $|l| = 1$ are illustrated in Figure 2. It is shown that when a Gaussian-shaped light beam is launched into the system, a phase singularity in the centroid of the beam starts to appear after the beam passing through the VR. The size of the dark area increases gradually over light propagation, giving rise to a hollow beam structure (see Figure 2). At the same time, the phase front of the optical-vortex beam starts to rotate, leading to a spiral phase profile with gradually-increasing bending radius. Simulations on light evolutions of optical-vortex beams with topological charges of $|l| = 2$ and 3 were also performed (see Section S1, Supporting Information).

## 4. Experimental results

In the experiments, we measured the ultrafast optical-vortex beams with topological charges of $l = -1$, 2 and 3 at the measurement position of ~0.7 m behind the VRs. The spatial-phase profiles of these vortex beams, measured using the tilted shearing-interferometer, are illustrated in the **Figure 3**. As shown in Figure 3a-c, at the same measurement position the diameter of the first ring of the hollow structure increases as the topological charge increases, agreeing well with the simulation results (see Figure S1, Supporting Information).

The tilted shearing-interference signal was recorded using the CCD camera, exhibiting clear interference patterns with high contrasts (see Figure 3d-f). With a single scan of the time delay between the two interfering light beams, a series of interferograms measured at the CCD camera can be written as:[35]

$$S(x,y,\tau)=\int\left|E(x,y,\tau)+E_d(x,y,t-\tau)\right|^2 dt \tag{1}$$

where $E(x,y,\tau)$ and $E_d(x,y,t-\tau)$ represent the electrical fields of green and pink (divergent) light beams shown in Figure 1a. The Fourier transform of $S(x,y,\tau)$ with respect to the delay time $\tau$ can be used to calculate directly the spectral phase difference between the two beams,[35] which can be expressed as $\Delta\varphi(x,y,\omega)=\varphi(x,y,\omega)-\varphi_d(x,y,\omega)$, where $\varphi(x,y,\omega)$ and $\varphi_d(x,y,\omega)$ are respectively the spectral phase distributions of the two beams. In this self-referencing scheme, these two phase distributions are closely correlated and can be expressed as:

$$\varphi_d(x,y,\omega)=\varphi(x_0'+\frac{x-x_0}{\beta},y_0'+\frac{y-y_0}{\beta},\omega)-\varphi_c(x,y,\omega) \tag{2}$$

where $(x_0,y_0)$ and $(x_0',y_0')$ represent the centroid positions of the pink and green beams, and $\beta$ is the beam-expansion rate due to the beam divergence. $\varphi_c$ is a spatially-varied phase due to the CM1, which is induced by the curvature of the mirror. While $\Delta\varphi(x,y,\omega)$ can be obtained using the meausured interferogram data, the precise values of $(x_0,y_0)$, $(x_0',y_0')$ and $\beta$ can be obtained in the experiment as (2140.2 μm, 350.6 μm), (1586.7 μm, 313.6 μm) and 1.4 for the case of $l=-1$. Then, we can retrieve the spectral phase distribution $\varphi(x,y,\omega)$ of the optical-vortex beam using an interative algorithm. Some detailed information of the retrievement process are presented in the Supporting Information, Section S2.

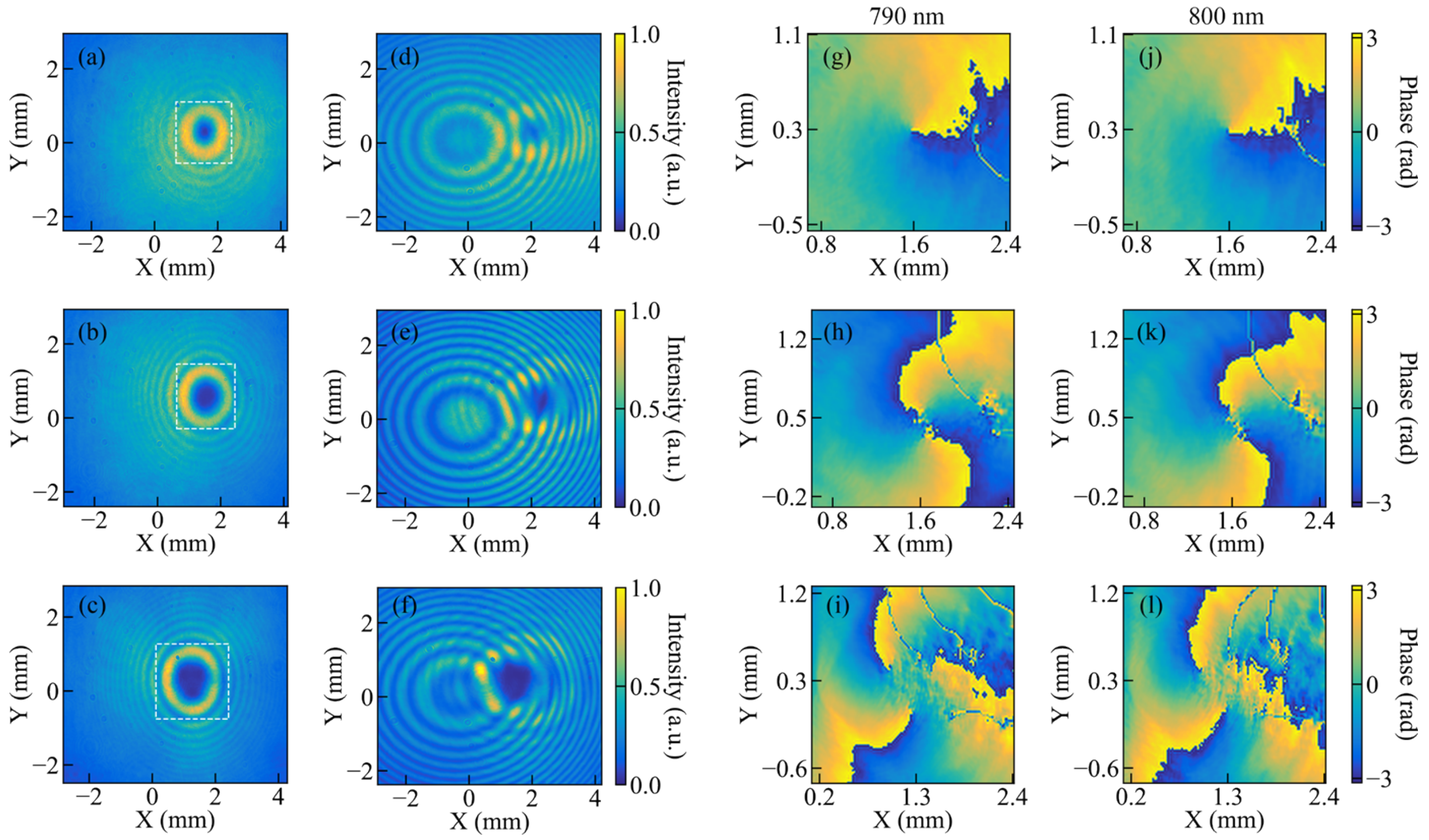

**Figure 3.** a-c) Measured intensity profiles of different optical-vortex beams ($l = -1$, 2 and 3). d-f) Typical results of tilted shearing-interference signal, measured using the CCD camera. g-l) Retrieved phase profiles of the corresponding optical-vortex beams at two wavelengths of 790 nm and 800 nm.

The retrieved spectral phase distributions of optical-vortex beams with different topological charges, at two wavelengths of 790 nm and 800 nm, are illustrated in Figure 3g-l, which show striking agreement with the simulated results shown in Figure 2 and Figure S1 in the Supporting Information. At the position of 0.7 m behind the VR, the measured phase profiles (see Figure 3g-l) possess phase singularities at the centroids of the beams, and the bending radius of the spiral structure increases as topological charge of the beam increases. It is noted that at topological charges of $l = 2$ and 3, some areas of phase ambiguity were observed near the phase singularities (see Figure 3h,k,i,l), which is mainly due to the fact that at high topological orders, the lower light intensities near the phase singularities lead to lower contrasts of the interference signal and therefore lower signal-to-noise ratios of the phase measurement at these areas.

Through a single scan of the interferometer arm, the pulse FROG trace at the sampling point of the vortex beam can also be measured using the spectrometer placed after the optical filter (see Figure 1a,b). A typical FROG trace, measured at $l = -1$, is illustrated in **Figure 4**a. A super-Gaussian filter was firstly used to extract useful information from the original data,[51] eliminating high-frequency noise as shown in Figure 4b. Then, the retrieved FROG trace was obtained using the extended ptychographic iterative engine (ePIE) algorithm,[52] see Figure 4c. The retrieved temporal intensity and phase profile (Figure 4d) yield a pulse duration of 53 fs together with an almost flat temporal phase curve. The optical spectrum retrieved from the FROG measurement exhibits good agreement with the pulse optical spectrum measured directly using the spectrometer (see Figure 4e).

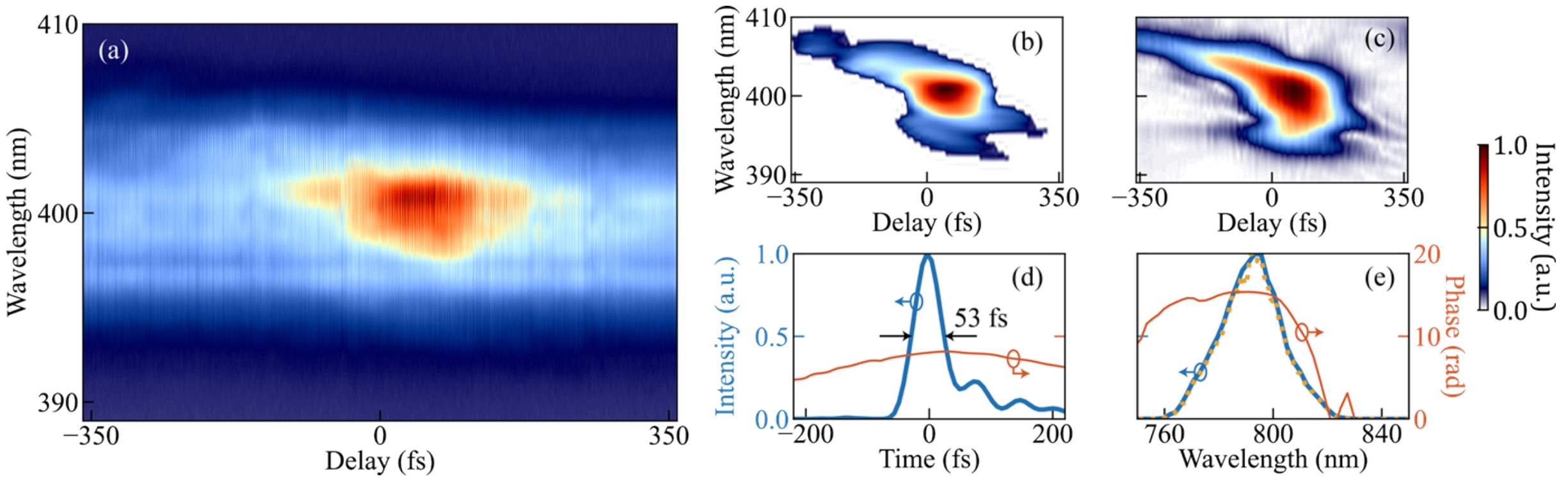

**Figure 4.** a) FROG trace recorded by the spectrometer for the case of $l = -1$. b) Filtered trace with high-frequency noise eliminated. c) Retrieved FROG trace using the ePIE algorithm. d,e) Temporal and spectral information of the pulse retrieved from the FROG trace. The yellow dashed curve in (e) gives the pulse spectrum, directly measured by the spectrometer.

With the relative spectral amplitude and phase distribution measured respectively by the CCD camera and tilted shearing interferometer, the spatio-temporal information over the whole spatial light beam can be obtained using the FROG trace at the sampling point as a reference. Therefore, the three-dimensional reconstruction of the ultrafast optical-vortex beams can be performed through a simple inverse Fourier transform procedure.[35] In the experiment, the complete three-dimensional electrical fields of ultrafast pulses with $l = -1$, 2 and 3 are reconstructed, and the results are illustrated in **Figure 5**. In order to make the vortex structure more obvious, in these plots we set the isosurface to be 40% of the peak amplitude of the electrical field and the carrier frequency was reduced to be 50% of the original one.

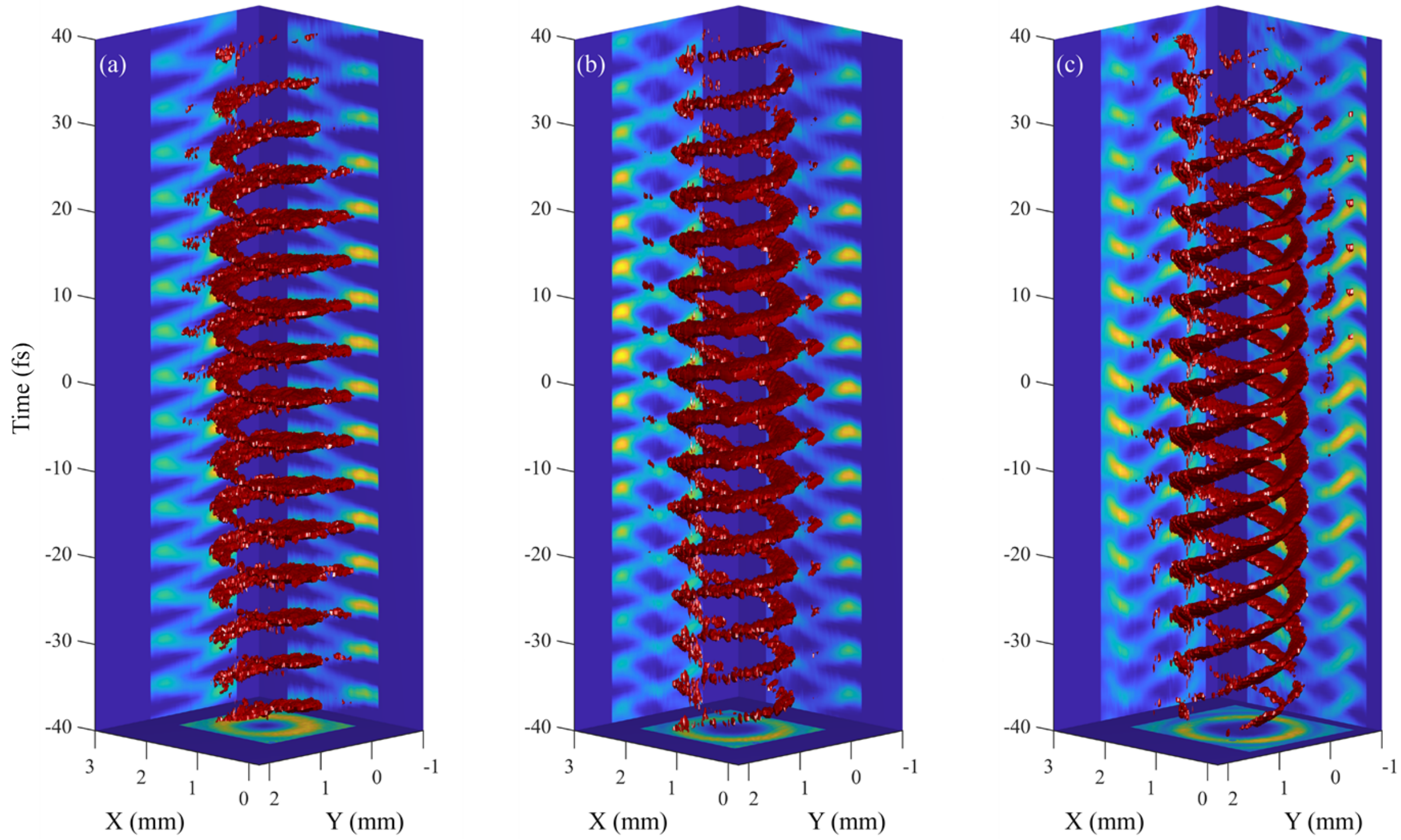

**Figure 5.** Reconstructed three-dimensional electrical fields of the optical-vortex pulses with respectively topological charges of (a) $l = -1$, (b) $l = 2$ and (c) $l = 3$.

## 5. Discussion and conclusion

The current set-up can merely be used to measure OAM-carrying ultrafast pulses with linear polarization states, which is mainly limited by the use of birefringent element and type-II phase-matching crystal in the system (see Figure 1a). However, this self-referencing technique could be extended further to characterize OAM-carrying pulses with more complicated polarization features. For simplest case, using a polarization beam splitter and two identical set-ups, simultaneous measurements of spatio-temporal structures of light beams at two orthogonal polarization states can be achieved. The remaining step is to build up the phase correlation between the two beams at orthogonal polarization states, which can be realized in principle through an additional interference measurement. Moreover, the application of this advanced measurement technique could be extended for measuring more complicated spatial-structured light beams,[53-56] and the tilted shearing-interferometer used in this set-up could also be modified according to different beam structures. Some further improvements on this TERMITES technique are also worth investigations. For example, through scanning the centroid off-set in the interferogram, the signal fading effect due to phase singularities could be largely suppressed, increasing further the signal-to-noise ratio of the measurement.

The use of type-II phase matching in the FROG measurement is critical for characterizing ultrafast optical-vortex beams. Comparing with conventional Gaussian-shaped light beam, optical-vortex beams with phase singularities generally have larger beam sizes.[4,5] In order to sample one particular point of the optical-vortex beam in the FROG measurement, the pulse-energy difference between two beams have to be carefully selected (997:3 in this experiment) so as to ensure almost-equal light intensities of the two beams at the sampling spot. If a conventional type-I crystal[37] was used in current set-up, the self-frequency-multiplication signal from the high-energy pulse would be much stronger than the sum-frequency one, leading to low signal-to-noise ratio of the FROG measurement. Some detailed results are presented in the Supporting Information, Section S3. When type-II crystal is used, the self-frequency-multiplication noise is strongly suppressed as a result of phase mismatch, ensuring a high signal-to-noise ratio of the measurement.

In conclusion, we demonstrated an advanced TERMITES scheme based on tilted shear-interference and type-II FROG measurement, being capable of characterizing ultrafast optical-vortex beam carrying OAMs in a self-referencing manner. The reconstructed three-dimensional electrical fields of ultrafast optical-vortex pulses with topological charges of $|l| = 1$, 2 and 3, exhibit striking agreement with numerical simulations, verifying the high performance of the scheme. The set-up reported here provides a novel means of characterizing spatially-structured ultrafast pulses, and may find many applications in OAM-related experiments and studies on nonlinear optics, ultrafast lasers and light-matter interactions.

**Acknowledgements**

This work was supported by the National Postdoctoral Program for Innovative Talents (BX2021328), the China Postdoctoral Science Foundation (2021M703325), the National Natural Science Foundation of China Youth Science Foundation Project (62205353), the Shanghai Science and Technology Innovation Action Plan (21ZR148270), the National High-level Talent Youth Project, the Zhangjiang Laboratory Construction and Operation Project (20DZ2210300), the National Natural Science Foundation of China (61925507), the Strategic Priority Research Program of the Chinese Academy of Sciences (XDB1603).

Supporting Information

# Self-referencing 3D characterization of ultrafast optical-vortex beams using tilted interference TERMITES technique

Jinyu Pan,[1,2,3,†] Yifei Chen,[1,†] Zhiyuan Huang,[1,*] Cheng Zhang,[3] Tiandao Chen,[1,2] Donghan Liu,[1,2] Ding Wang,[1] Meng Pang,[1,3,*] and Yuxin Leng[1,3,*]

[1]*State Key Laboratory of High Field Laser Physics and CAS Center for Excellence in Ultra-intense Laser Science, Shanghai Institute of Optics and Fine Mechanics, Chinese Academy of Sciences, Shanghai 201800, China*

[2]*Center of Materials Science and Optoelectronics Engineering, University of Chinese Academy of Sciences, Beijing 100049, China*

[3]*Hangzhou Institute for Advanced Study, University of Chinese Academy of Sciences, Hangzhou 310024, China*

[*]*E-mail: huangzhiyuan@siom.ac.cn; pangmeng@siom.ac.cn; lengyuxin@siom.ac.cn*

[†]*These authors contributed equally to this work.*

**S1. Vortex retarders and simulated evolutions of optical-vortex beams**

In the experiments we used vortex retarders for generating optical vortex beams. In the liquid crystal polymer layer of the vortex retarder, the fast-axis orientation of the liquid crystal molecules continuously changes along with the angular direction, expressed as:[1]

$$\phi = \frac{l}{2}\theta \tag{S1}$$

where $l$ is the order of the vortex retarder, $\phi$ is the fast-axis orientation at the specific position of $\theta$, see **Figure S1**a-c. When a circularly-polarized light beam passes through the vortex retarder, it will obtain a helical phase front so as to form an optical-vortex beam. In addition, we can convert the circular polarization state of the vortex beam into the linear polarization state using a wire grid polarizer.

When a helical phase front is loaded on a Gaussian-shaped light beam, the electrical field evolution of the light beam in free space can be described using the Collins diffraction integral formula, written as:[2,3]

$$E(\rho,\theta,z,\omega)=\sqrt{2\pi}T_0\exp[-\frac{T_0^2}{2}(\omega-\omega_0)^2]\frac{(-i)^{|l|+1}z_R^2}{(z-iz_R)^{3/2}}(\frac{\pi\rho^2}{4zw_0^2})\exp[i(kz+l\theta)]$$
$$\times\exp(\frac{ik\rho^2}{2z})\exp[-\frac{z_R^2\rho^2/w_0^2}{2z(z-iz_R)}]\times[I_{\frac{|l|-1}{2}}(\frac{z_R^2\rho^2/w_0^2}{2z(z-iz_R)})-I_{\frac{|l|+1}{2}}(\frac{z_R^2\rho^2/w_0^2}{2z(z-iz_R)})] \quad \text{(S2)}$$

where $\rho$ is the radial coordinate, $\theta$ is the azimuthal angle, and $z$ is the propagation distance. $T_0$ and $w_0$ are respectively the initial pulse duration and the beam width. $z_R = k\omega_0^2/2$ is the Rayleigh distance of the Gaussian beam, where $k$ is the wave vector and $\omega_0$ is the central angular frequency. $l$ and $I_m(\bullet)$ represent the corresponding topological charge and the modified Bessel function.

The spatial light intensity can be obtained by integrating the square of the complete electrical field of the optical-vortex beam over $\omega$:

$$I(\rho,\theta,z)=\int\left|E(\rho,\theta,z,\omega)\right|^2 d\omega \quad \text{(S3)}$$

The simulated intensity evolutions of optical-vortex beams with topological charges of $l = -1$, 2 and 3 are illustrated in Figure S1d-f, and the corresponding intensity distributions at the propagation position of 0.7 m are plotted in Figure S1g-i. It is shown that the size of the central dark region of the beam increases with the propagation distance, and a larger mode area is obtained at a higher topological charge. In addition, we can obtain the phase profile from the imaginary part of the electrical field of the optical-vortex beam. In Figure S1j-l, the corresponding phase profiles are plotted at the wavelength of 790 nm. It can be seen that as the topological charge increases, the spiral phase profile of the optical-vortex beam exhibits a gradually-increasing bending radius.

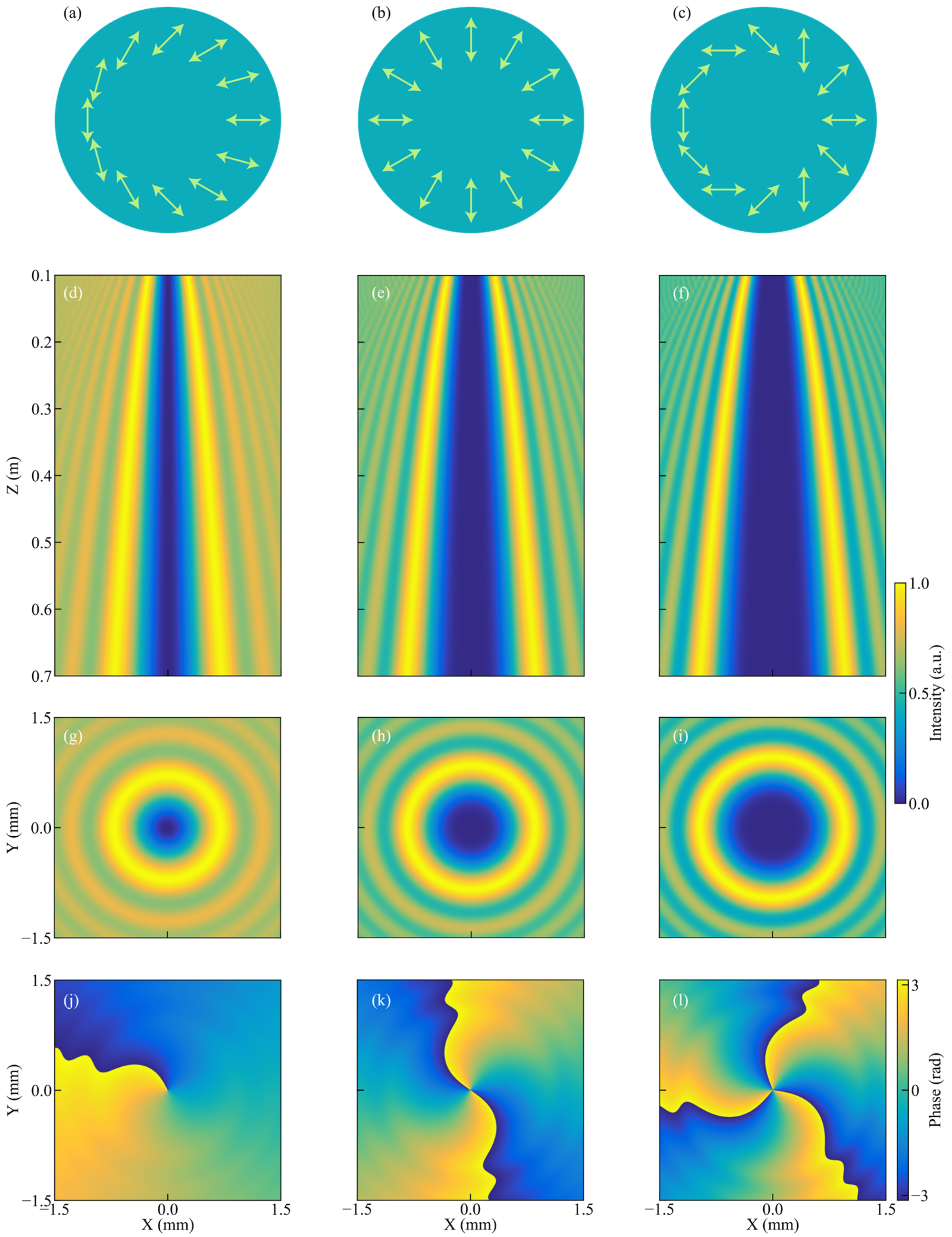

**Figure S1.** a-c) Vortex retarders with $|l|$ = 1, 2 and 3 have different types of fast-axis orientations. d-f) Simulated intensity evolutions of optical-vortex beams with topological charges of $l$ = −1, 2 and 3. g-i) Simulated intensity distributions of these optical-vortex beams at the propagation position of 0.7 m. j-l) The corresponding phase profiles at the wavelength of 790 nm.

## S2. Data processing for retrieving spectral phase distribution

The first-order cross-correlation function of the green and pink light beams in Figure 1 can be extracted from the interferogram signal, expressed as:[4-6]

$$s(x,y,\tau) = \int dt E(x,y,t) E_d^*(x,y,t-\tau) \tag{S4}$$

where $E(x,y,\tau)$ and $E_d(x,y,t-\tau)$ represent the electrical fields of green and pink (divergent) light beams, $E_d^*(x,y,t-\tau)$ is the complex conjugate of the electrical field $E_d(x,y,t-\tau)$. The phase of the Fourier transform of the cross-correlation function reflects the phase difference between the two beams, which can be expressed as:

$$\Delta\varphi(x,y,\omega) = \arg\left[\tilde{s}(x,y,\omega)\right] = \varphi(x,y,\omega) - \varphi_d(x,y,\omega) \tag{S5}$$

where $\varphi(x,y,\omega)$ and $\varphi_d(x,y,\omega)$ are the spectral phases of the green and pink (divergent) beams, $\tilde{s}(x,y,\omega)$ is the Fourier transform of the cross-correlation function, and $\varphi_d(x,y,\omega)$ is composed of two parts as:

$$\varphi_d(x,y,\omega) = \varphi_R(x,y,\omega) - \varphi_c(x,y,\omega) \tag{S6}$$

where $\varphi_R(x,y,\omega)$ represents the phase distribution induced by the coordinate transformation of the tilted convex mirror, and $\varphi_c(x,y,\omega)$ is a spatially-varied phase profile induced by the curvature of the convex mirror.

In the experiment, in order to remove $\varphi_c(x,y,\omega)$ we performed a second-order polynomial fit on $\arg\left[\tilde{s}(x,y,\omega_0)\right]$ at the central frequency, which was commonly used in conventional TERMITES techniques.[4,5] In this fitting procedure, the phase change caused by the rotation angle of the convex mirror was included, which can be expressed as:[4]

$$\arg\left[\tilde{s}(x,y,\omega_0)\right] = a_0 + a_1x + a_2y + a_3xy + a_4x^2 + a_5y^2 + O(x^2) + O(y^2) \tag{S7}$$

where $a_0$, $a_1$, $a_2$, $a_3$, $a_4$, and $a_5$ are the free parameters. Neglecting the higher order terms, we found that the fitting results using Equation (S7) is extremely accurate, and the phase $\varphi_c(x,y,\omega_0)$ can be written as:

$$\varphi_c(x,y,\omega_0) = a_1x + a_2y + a_3xy + a_4x^2 + a_5y^2 \tag{S8}$$

We can then calculate the spatially-varied phase $\varphi_c(x,y,\omega)$ at all the other frequencies, which can be expressed as:[4]

$$\varphi_c(x,y,\omega) = \frac{\omega}{\omega_0}\varphi_c(x,y,\omega_0) \tag{S9}$$

Subtracting the phase $\varphi_c(x,y,\omega)$ out of $\arg\left[\tilde{s}(x,y,\omega)\right]$ and using Equation (S5) and (S6), the phase $\Delta\varphi'(x,y,\omega)$ can be defined as:

$$\Delta\varphi'(x,y,\omega) = \arg\left[\tilde{s}(x,y,\omega)\right] - \varphi_c(x,y,\omega) \\ = \varphi(x,y,\omega) - \varphi_R(x,y,\omega) \tag{S10}$$

The method we used to extract $\varphi(x,y,\omega)$ from $\Delta\varphi'(x,y,\omega)$ is a simple iterative algorithm which was commonly used in conventional TERMITES techniques.[4,5] Since a tilted convex mirror was used in the set-up, the expression for $\varphi_R(x,y,\omega)$ needs to be slightly modified. **Figure S2** shows the 1D geometry of the device, where $x_c$ is the coordinate of the interference center on the CCD camera, $x_0$ and $x_0'$ represent the centroid positions of the pink and green beams in Figure 1. $x$ is the coordinate of the pink beam, and $x'$ is the coordinate of the green beam and $F'$ is the virtual image point.

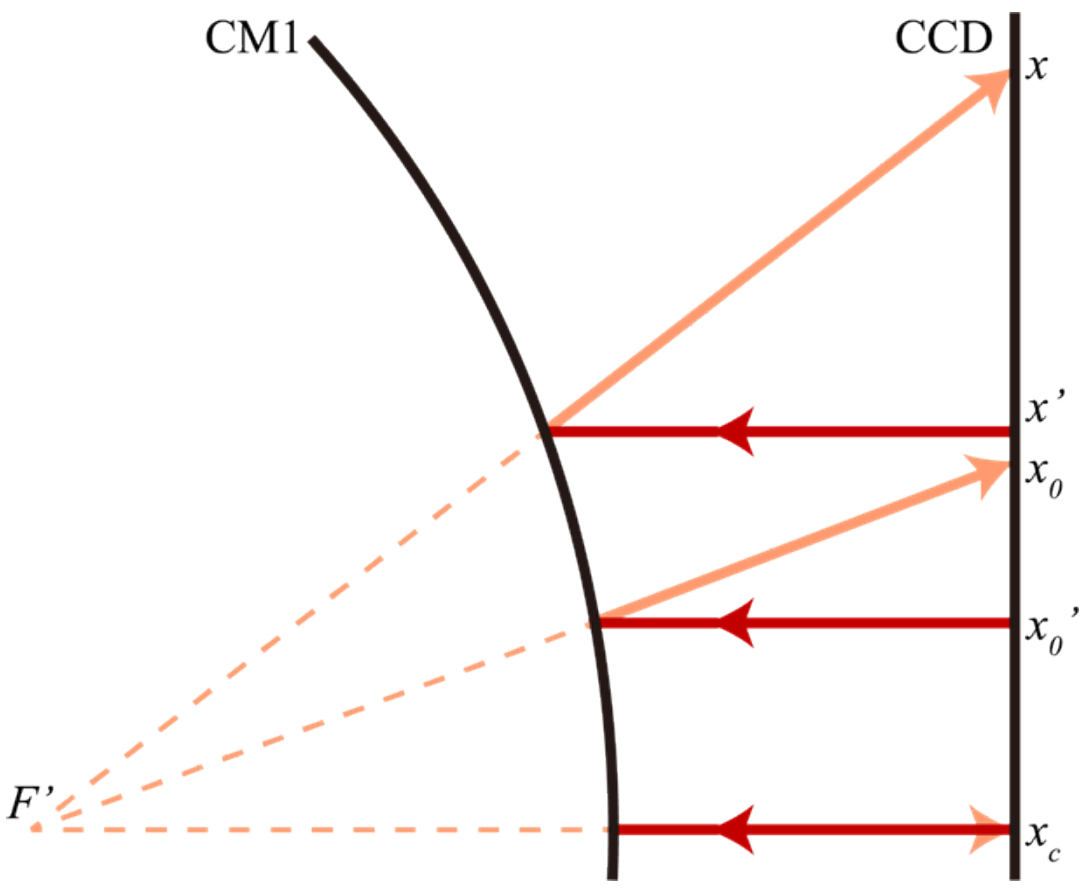

**Figure S2.** 1D geometry of the device.

As illustrated in Figure S2, the relation between $x$ and $x'$ can be expressed as:

$$x' = x_0' + \frac{x - x_0}{\beta_x} \tag{S11}$$

$$\beta_x = \frac{\sqrt{L^2 + (x_0 - x_c)^2}}{f} \tag{S12}$$

where $\beta_x$ is the beam-expansion rate due to the beam divergence, $f$ is the focal length of the convex mirror, and $L$ is the distance from the virtual image point of the convex mirror to the CCD. It should be noted that with the assumption of $|x_0 - x_c| << L$, $\beta_x$ can be written as:

$$\beta_x \approx \frac{L}{f} = \beta \tag{S13}$$

This relation also applies to the y dimension. Thus, the relation between $\varphi_R(x,y,\omega)$ and $\varphi(x,y,\omega)$ can be written as:

$$\varphi_R(x,y,\omega) = \varphi(x_0' + \frac{x - x_0}{\beta}, y_0' + \frac{y - y_0}{\beta}, \omega) \tag{S14}$$

In order to increase the measurement accuracy, we used the retrieved phase profiles around $x_0'$ to reconstruct the 3D electrical fields (see Figure 3 and Figure 5 in the main text). When the small angle of the convex mirror is removed, $x_0$, $x_0'$, and $x_c$ coincides at the same point, then Equation (S11) becomes:[4]

$$x' = \frac{x}{\beta} \tag{S15}$$

**S3. Comparison of FROG measurements using type-I and type-II phase-matching crystals**

In our set-up, a collinear FROG layout was used for measuring the temporal shape of the ultrafast pulse. The short-wavelength (~400 nm) signal generated from the BBO crystal were filtered out using an optical filter and then recorded using a spectrometer. Besides sum-frequency signal between the two beams, the filtered short-wavelength signal also includes the self-frequency multiplication signal from each of the beams, being a portion of the noise background in our measurements. For accurate FROG-trace retrieving, the self-frequency-multiplication signal should be suppressed in order to obtain a high signal-to-noise ratio of the sum-frequency signal. In the experiment, due to the convex mirror (see Figure 1a) the divergent pink beam had a large focal spot in the BBO crystal, while the green beam was tightly focused in the crystal with a small spot size, working as the sampling point of the pink beam. In order to maximize the sum-frequency signal, the total pulse energy difference between the pink and green beams was adjusted to be quite large (997:3 in the experiment) so as to obtain almost equal light intensities of the two beams at the sampling point. If type-I phase-matching crystal was used in this FROG measurement, this large energy difference would result in very strong self-frequency-multiplication signal generated from the high-energy pink beam, overwhelming the sum-frequency signal.

To suppress self-frequency-multiplication signal and enhance the signal-to-noise ratio of sum-frequency signal, we used in the experiment a type-II BBO crystal. In the set-up, a birefringent quartz crystal (BQC) was used to generate two light beams with orthogonal polarization states. As described in the main text, through carefully selecting the orientation of the BQC principle axis and the BQC thickness, the pink and green light beams can be generated at orthogonal polarization states and the two beams were focused in the BBO crystal, enabling a type-II phase-matching process of the sum-frequency generation. This type-II phase matching efficiently suppressed the generation of self-frequency-multiplication signal from each of the two beams, enhancing the signal-to-noise ratio of the FROG measurement.

**Figure S3** illustrates the enhancement of signal-to-noise ratio when type-II crystal was used in the FROG measurement for the case of $l = -1$ ultrafast optical-vortex-beam characterization. As shown in Figure S3a, when the type-I BBO crystal was used for the FROG measurement, the measured sum-frequency signal recorded by the spectrometer exhibited a poor signal-to-noise ratio, with quite strong background noise (self-frequency-multiplication signal). As shown in Figure S3b, in this case the super-Gaussian filter cannot efficiently remove the high-frequency noise and the filtered FROG trace is still quite noisy, degrading the quality of FROG-trace retrieving (see Figure S3c,d). In contrast, when type-II BBO crystal was used in the set-up, it is found that the measured FROG trace exhibits significantly-improved signal-to-noise ratio, as shown in Figure S3e-h. Note that the data plotted in Figure S3e is exactly the same as in Figure 4a of the main text, and the results after the super-Gaussian filtering is plotted in Figure S3f.

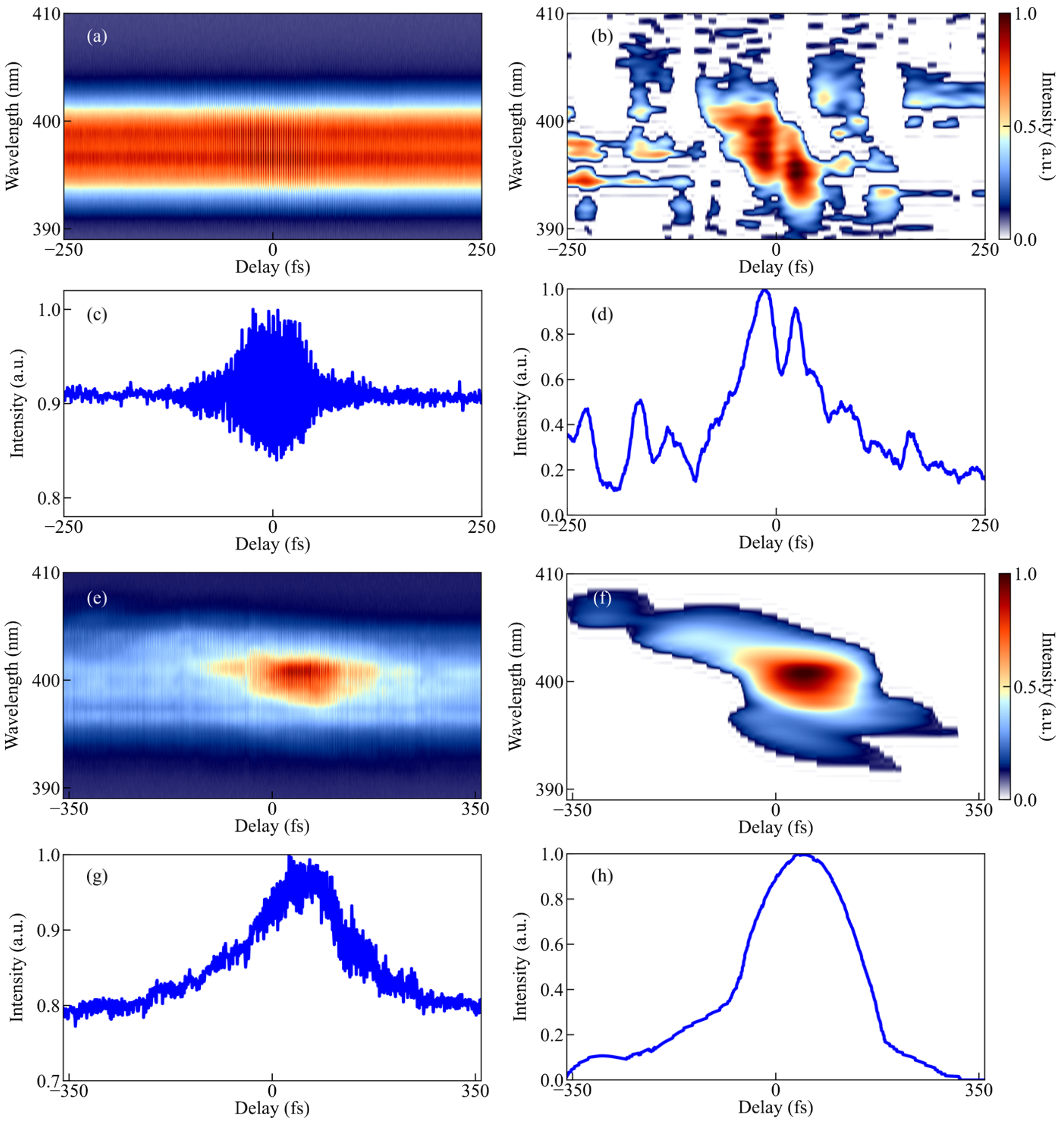

**Figure S3.** The measured FROG results using the type-I (a)-(d) and type-II (e)-(h) phase matching. a,e) Measured FROG traces recorded by the spectrometer for the case of $l = -1$. b,f) Filtered results using a super-Gaussian filter. c,d,g,h) Correlation traces corresponding to (a), (b), (e) and (f).